\documentclass[english,10pt]{iopart}
\usepackage{lmodern}
\usepackage[T1]{fontenc}
\usepackage[latin9]{inputenc}
\usepackage{verbatim}
\usepackage{textcomp}
\usepackage{bm}
\usepackage{graphicx}
\usepackage{color}
\usepackage{cite} 

\makeatletter
\@ifundefined{textcolor}{}
{%
 \definecolor{BLACK}{gray}{0}
 \definecolor{WHITE}{gray}{1}
 \definecolor{RED}{rgb}{1,0,0}
 \definecolor{GREEN}{rgb}{0,1,0}
 \definecolor{BLUE}{rgb}{0,0,1}
 \definecolor{CYAN}{cmyk}{1,0,0,0}
 \definecolor{MAGENTA}{cmyk}{0,1,0,0}
 \definecolor{YELLOW}{cmyk}{0,0,1,0}
 }

\makeatother

\usepackage{babel}

\begin{document}

\title[Absence of strong magnetic fluctuations in FeP-based systems]{Absence of strong magnetic fluctuations in FeP-based systems LaFePO and Sr$_{2}$ScO$_{3}$FeP}

\author{A. E. Taylor$^1$, R. A. Ewings$^2$, T. G. Perring$^{2,3}$, D. R. Parker$^4$, J. Ollivier$^5$, S. J. Clarke$^4$ and A. T. Boothroyd$^1$}
\address{$^1$ Department of Physics, University of Oxford, Clarendon Laboratory,
Parks Road, Oxford, OX1 3PU, United Kingdom}
\address{$^2$ ISIS Facility, Rutherford Appleton Laboratory, STFC, Chilton, Didcot,
Oxon, OX11 0QX, United Kingdom}
\address{$^3$ London Centre for Nanotechnology, University College London, London, WC1H 0AH, United Kingdom}
\address{$^4$ Department of Chemistry, University of Oxford, Inorganic Chemistry Laboratory, South Parks Road, Oxford OX1 3QR, United Kingdom}
\address{$^5$ Institut Laue--Langevin, 6 rue Jules Horowitz, BP 156, F-38042, Grenoble Cedex 9, France}
\ead{a.taylor1@physics.ox.ac.uk}
\ead{a.boothroyd@physics.ox.ac.uk}

\begin{abstract}
We report neutron inelastic scattering measurements on polycrystalline LaFePO and Sr$_{2}$ScO$_{3}$FeP, two members of iron-phosphide families of superconductors. No evidence is found for any magnetic fluctuations in the spectrum of either material in the energy and wave vector ranges probed. Special attention is paid to the wave vector at which spin density wave-like fluctuations are seen in other iron-based superconductors. We estimate that the magnetic signal, if present, is at least a factor of four (Sr$_2$ScO$_3$FeP) or seven (LaFePO) smaller than in the related iron arsenide and chalcogenide superconductors. These results suggest that magnetic fluctuations are not as influential to the electronic properties of the iron phosphide systems as they are in other iron-based superconductors.
\end{abstract}


\maketitle


\section{Introduction}

It is widely believed that magnetic fluctuations are involved in the superconducting mechanism of the iron-based superconductors, but there is currently no complete understanding of the microscopic origin of magnetism or of its detailed relationship with superconductivity in these materials~\cite{mazin_pairing_2009,hirschfeld_gap_2011,stewart_superconductivity_2011,dai_magnetism_2012}.
Strong magnetic fluctuations related to those found in the non-superconducting spin density wave (SDW) parent phases have been observed in nearly all families of iron  arsenide and chalcogenide superconductors by inelastic neutron scattering (INS)~\cite{stewart_superconductivity_2011,dai_magnetism_2012,johnston_puzzle_2010,lumsden_magnetism_2010}.
These fluctuations are found to persist across the phase diagram, including for systems which show no SDW ordering.
In superconducting samples the INS spectrum contains a particularly prominent feature known as the spin resonance. This is a peak localized both in momentum and in energy which grows strongly in intensity at temperatures below $T_{\mathrm{c}}$, and is a key piece of evidence linking magnetic fluctuations with superconductivity~\cite{christianson_unconventional_2008}.

The iron phosphide systems differ in several respects from their arsenide counterparts. They generally have lower $T_{\mathrm{c}}$ values, and the stoichiometric compounds do not undergo magnetic or structural phase transitions~\cite{kamihara_iron-based_2006,mcqueen_intrinsic_2008,ogino_superconductivity_2009,deng_new_2009,mydeen_temperature-pressure_2010}. Evidence has been found for a nodal superconducting order parameter in LaFePO~\cite{hicks_evidence_2009,fletcher_evidence_2009,sutherland_low-energy_2012},
Sr$_{2}$ScO$_{3}$FeP~\cite{yates_evidence_2010} and LiFeP~\cite{hashimoto_nodal_2012}, in contrast to many of the iron arsenide and selenide superconductors. The difference in gap structure has been explained by the absence of a Fermi surface hole pocket centred on the $M$ point in the Brillouin zone in the phosphides~\cite{thomale_mechanism_2011}. As far as spin fluctuations are concerned there is conflicting evidence. Optical and charge transport studies have concluded that electron correlations are significantly weaker in the phosphides than in the arsenides~\cite{qazilbash_electronic_2009,kasahara_contrasts_2012}, whereas measurements of the Fermi surface indicate that correlations could be of similar strength in the two systems~\cite{coldea_fermi_2008}. Furthermore, evidence of spin correlations in phosphides has been reported in NMR and $\mu$SR studies~\cite{nakai_spin_2008,carlo_static_2009}.

These differences raise the possibility that the pairing mechanism in the iron phosphides might not be the same as that in the higher $T_{\mathrm{c}}$ iron-based superconductors. In particular, the role of magnetic fluctuations in the pairing mechanism of the iron phosphides remains unresolved, and this provides a strong incentive to obtain experimental information on the momentum-resolved magnetic fluctuation spectrum. Such information is so far lacking for the phosphides.

Here we present results of INS measurements to search for magnetic fluctuations in two phosphides, LaFePO and Sr$_{2}$ScO$_{3}$FeP. LaFePO has a relatively low $T_{\mathrm{c}}$ of 4.5\,K, but is of interest because of evidence that
it might be close to a SDW instability driven by Fermi surface nesting~\cite{coldea_fermi_2008,thomale_mechanism_2011}. Sr$_{2}$ScO$_{3}$FeP was chosen because it was reported to show the highest $T_{\mathrm{c}}$($\approx17\,$K) among the known phosphide superconductors~\cite{ogino_superconductivity_2009}.

At present, large single crystals of iron phosphides are not available, so our measurements were performed on polycrystalline samples. We searched for magnetic fluctuations with energies $E$ in the range from 0.5 to 20\,meV, and explored a large range of wave vector $Q = |{\bf Q}|$ space.
We paid particular attention to $Q \sim 1.2\,\mathrm{\AA}^{-1}$ corresponding to the wave vector ${\bf Q}_{\rm SDW}$ of the SDW order found in the parent phases of many iron-based superconductors (the in-plane component of ${\bf Q}_{\rm SDW}$ is $\left(0.5,0\right)$ when expressed in reciprocal lattice units of the one-Fe unit cell). We looked closely at energies in the region $E \sim 5k_{\rm B}T_{\rm c}$ where the spin resonance has been observed in iron-based superconductors. Despite performing careful measurements above and below $T_{\rm c}$, we found no signal attributable to magnetic fluctuations of any kind in either system. The results suggest that magnetic fluctuations might not be as important for superconductivity in the iron phosphide systems as they are in other iron-based superconductors.


\section{Experimental Methods\label{sec:Experimental-Methods}}

\subsection*{Synthesis}

Polycrystalline samples were prepared by the following solid-state
reaction routes. All manipulations were carried out in an argon-filled glovebox with a combined O$_{2}$ and H$_{2}$O
content of less than 5$\,$parts per million. The silica ampoules
were heated to 1000$\,^{\circ}$C under vacuum to remove any trapped
moisture before use.

\paragraph*{LaFePO:}

LaFePO was prepared based on a combination of the reaction routes
reported by Kamihara \emph{et al.}~\cite{kamihara_iron-based_2006}
and McQueen \emph{et al.}~\cite{mcqueen_intrinsic_2008}. La$_{2}$O$_{3}$
powder was prepared by dehydrating commercial powder (Alfa 99.99\,\%) at
600$\,^{\circ}$C for 10$\,$hours. Fresh La metal (Alfa 99.9\,\%) filings, red P
(Alfa 99.9999\,\%) ground to a powder, and Fe powder (Alfa 99.998\,\%) were mixed
in the ratio 1:3:3 and placed in an alumina crucible in an evacuated
silica ampoule. The elements were heated at 0.5$\,^{\circ}$C/min to 700$\,^{\circ}$C
and held for $\sim$12$\,$hours, to produce a mixture of the compounds
LaP, FeP and Fe$_{2}$P. This mixture and La$_{2}$O$_{3}$ were then
ground together to a very fine powder which was placed in an alumina
crucible; another crucible containing a LaFePO getter (synthesised
in a similar way) was placed on top~\cite{mcqueen_intrinsic_2008}. This was all sealed in a silica
ampoule under 200$\,$mbar of high purity argon gas in order to prevent
collapse of the silica upon heating at close to the softening temperature of silica. This was heated at 1$\,^{\circ}$C/min
to 1250$\,^{\circ}$C and held for 24$\,$hours. The final product was removed and ground to a fine powder.

\paragraph*{LaZnPO:}

LaZnPO was prepared starting from fresh La metal filings, red P
powder and ZnO (Alfa 99.99\,\%) powder, following the route previously described
by Kayanuma \emph{et al.}~\cite{kayanuma_apparent_2007}

\paragraph*{Sr$_{2}$ScO$_{3}$FeP:}

Sr$_{2}$ScO$_{3}$FeP was prepared according to the following route.
SrO powder was prepared by the thermal decomposition of SrCO$_{3}$
(Alfa 99.99\,\%) by heating to 850\,$^{\circ}$C and holding for 16 hours, then 1100\,$^{\circ}$C for 4 hours, under dynamic vacuum. Sr metal (Aldrich 99\,\%) was sublimed under high vacuum
at 850 \textdegree{}C prior to use. Red P, ground to a powder,
and small pieces of Sr were placed in a Nb tube in the ratio 1:1. The
Nb tube was then sealed by welding in an arc furnace under one atmosphere of argon,
and then sealed in a protective evacuated silica tube. This was heated
at 2$\,^{\circ}$C/min to 800$\,^{\circ}$and held for 3$\,$days
and the resulting mixture was ground to a fine powder.
This powder was mixed with SrO, Sc$_{2}$O$_{3}$ powder (99.99\,\%),
Fe powder, and Fe$_{2}$O$_{3}$ powder (Alfa 99.998\,\%) according
to the stoichiometry Sr$_{2}$ScO$_{3}$FeP. After homogenization, this
powder was pelletized and placed in an alumina crucible, then sealed
in an evacuated silica ampoule. This was heated at 2$\,^{\circ}$C/min
to 1200$\,^{\circ}$C and held for 10$\,$hours. The final product was removed
and ground to a fine powder.

\subsection*{Characterization}

Room temperature x-ray powder diffraction (XRPD) was used to assess
the phase purity of all the products prepared as described above.
Measurements were made with a PANalytical X\textquoteright{}Pert
PRO diffractometer operating in Bragg\textendash{}Brentano geometry
with monochromatic Cu K$_{\alpha1}$ radiation and a multiangle X\textquoteright{}Celerator
detector. Structural refinements against XRPD data were carried out
using the Rietveld refinement package GSAS+EXPGUI~\cite{larson_general_1994,toby_expgui_2001}.
DC susceptibility measurements were performed from 2$\,$K to 300$\,$K
in a measuring field of 50$\,$Oe using a Quantum Design MPMS-XL SQUID
magnetometer.

\paragraph*{LaFePO:}

\begin{figure*}
\begin{centering}
\includegraphics[width=0.95\textwidth]{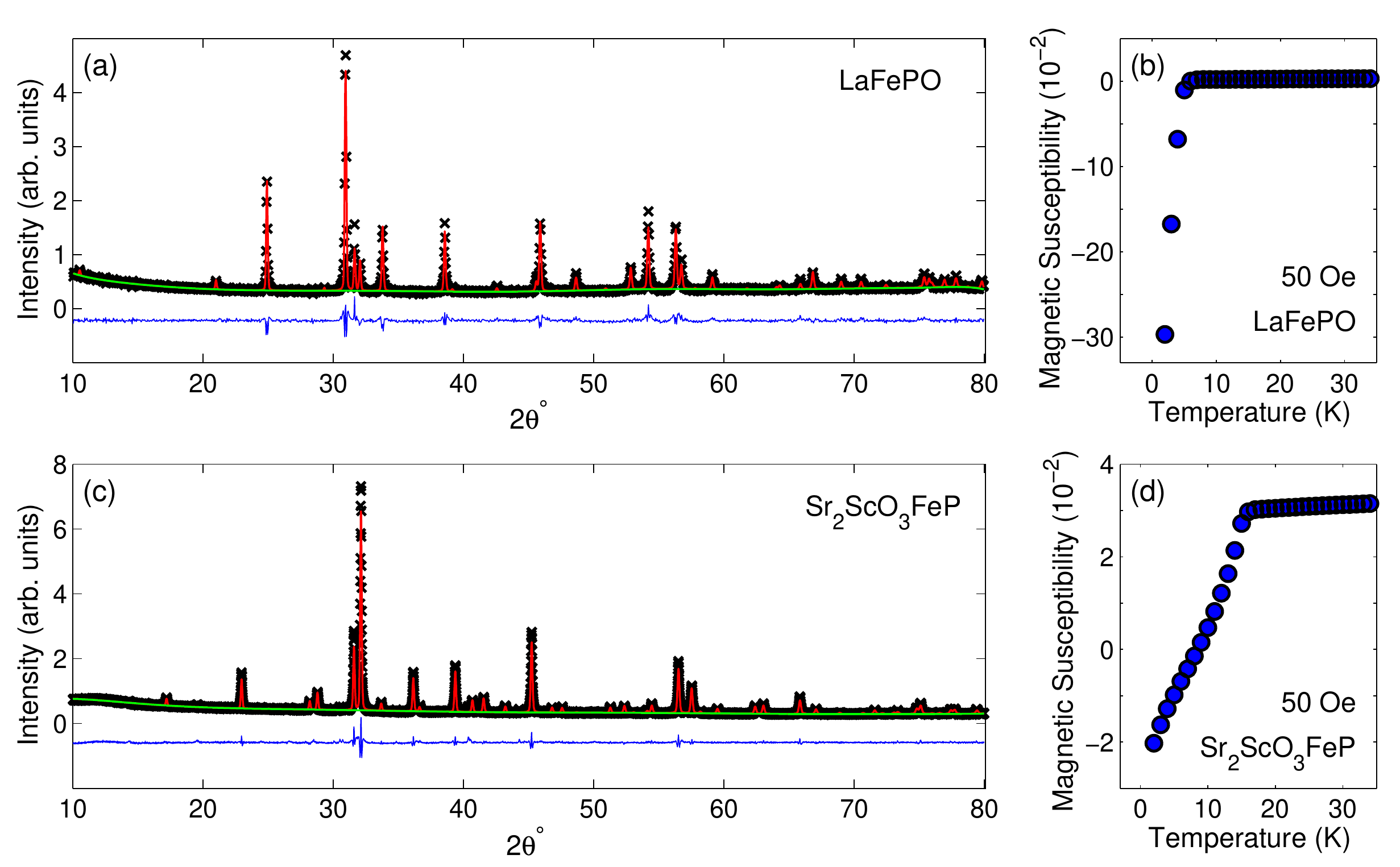}
\caption{\label{fig:Both_xray_SQUID} Characterization of LaFePO, (a) and (b), and Sr$_2$ScO$_3$FeP, (c) and (d). (a) and (c) show Rietveld
refinements against x-ray powder diffraction data (black crosses) taken at room temperature, the calculated background
and the difference between calculation and data are also shown. (b) and (d) show the results of magnetic susceptibility measurements made on portions of the samples
in an applied field of 50\,Oe under zero-field-cooled conditions. Susceptibility is expressed in SI units, with $-1$ corresponding to perfect diamagnetism.}
\end{centering}
\end{figure*}

A typical result of the characterization of LaFePO is shown in figures~\ref{fig:Both_xray_SQUID}(a) and~(b). Figure~\ref{fig:Both_xray_SQUID}(a)
shows XRPD data together with the calculated pattern based on the results of
the Rietveld refinement. We find lattice parameters $a=b=3.9646(2)\,\mathrm{\AA}$
and $c=8.5187(5)\,\mathrm{\AA}$ for space group $P4/nmm$, where the error is the estimated standard deviation calculated by GSAS. The standard deviation of lattice parameter values measured on several samples was $\sigma(a) = 0.0003\,\mathrm{\AA}$ and $\sigma(c) = 0.001\,\mathrm{\AA}$.  Figure~\ref{fig:Both_xray_SQUID}(b) shows a zero-field-cooled magnetic susceptibility measurement on the
same sample. We confirm $T_{\mathrm{{c}}}\approx4.5\,$K
and we estimate the overall superconducting volume fraction of the
sample used for the neutron scattering experiments to be \textasciitilde{}20$\,$\% at 2\,K,
consistent with previous reports~\cite{kamihara_iron-based_2006,kamihara_electromagnetic_2008}.

\paragraph*{LaZnPO:}

LaZnPOis isostructural with LaFePO and was synthesized to use as a non-magnetic background for comparison
to LaFePO in the neutron scattering experiments. XRPD revealed a phase
pure sample and the refined structure gave lattice parameters $a=b=4.04203(3)\,\mathrm{\AA}$
and $c=8.90626(9)\,\mathrm{\AA}$ with space group $P4/nmm$.

\paragraph*{Sr$_{2}$ScO$_{3}$FeP:}

A typical result of the characterization of Sr$_{2}$ScO$_{3}$FeP
is shown in figures~\ref{fig:Both_xray_SQUID}(c) and~(d). Figure~\ref{fig:Both_xray_SQUID}(c)
shows XRPD data along with the calculated pattern based on the results
of the Rietveld refinement for Sr$_{2}$ScO$_{3}$FeP with space group
$P4/nmm$. Unlike the previous report on this material by
Ogino \emph{et al.}~\cite{ogino_superconductivity_2009}, in which
they found Sr$_{2}$ScO$_{3}$FeP and SrFe$_{2}$P$_{2}$ to coexist
in the ratio 9:1, figure~\ref{fig:Both_xray_SQUID}(c) shows that
the Sr$_{2}$ScO$_{3}$FeP phase alone describes our data very well.
We find lattice parameters  $a=b=4.0148(3)\,\mathrm{\AA}$ with $\sigma(a) = 0.002\,\mathrm{\AA}$ and $c=15.551(2)\,\mathrm{\AA}$ with $\sigma(c) = 0.02\,\mathrm{\AA}$, where errors are determined as described above for LaFePO. These lattice parameters are similar within the error to those previously reported~\cite{ogino_superconductivity_2009}.
Magnetic susceptibility measurements shown in figure~\ref{fig:Both_xray_SQUID}(d)
established the onset of superconductivity occurs at $T_{\mathrm{{c}}}\approx15\,$K,
but it appears to be related to a superconducting volume fraction of only a few percent. By contrast, the sample reported by Ogino {\it et al.}~\cite{ogino_superconductivity_2009} that contained small amounts of impurity phases was a bulk superconductor. This suggests, therefore, that the superconducting phase is most likely a slightly off-stoichiometric form of Sr$_2$ScO$_3$FeP, and that Sr$_{2}$ScO$_{3}$FeP is close to a superconducting instability  but may not be an intrinsic bulk superconductor. Further investigation on this point is beyond the scope of this article.
%
%

\subsection*{Inelastic Neutron Scattering}

The inelastic neutron scattering experiments were performed on the
time-of-flight chopper spectrometers MERLIN at the ISIS Facility~\cite{bewley_merlin_2006},
and IN5 at the Institut Laue--Langevin~\cite{ollivier_in5_2011}, both
of which have large, position-sensitive detector arrays. These instruments
allow measurement of vast regions of $(Q,E)$ space, which is advantageous when searching throughout
the Brillouin zone for evidence of magnetic excitations. For each measurement the powder was sealed in
an annulus around the edge of a cylindrical aluminium can. 10.6$\,$g of LaFePO, 8.8$\,$g of Sr$_{2}$ScO$_{3}$FeP and 7.9$\,$g
of LaZnPO powder were measured. On MERLIN, LaFePO and Sr$_{2}$ScO$_{3}$FeP
were measured with incident neutron energies $E_{\mathrm{{i}}}=25$
and $50\,$meV at temperatures of $T=6$ and $20\,$K. On IN5, LaFePO
and LaZnPO were measured with incident neutron energies $E_{\mathrm{{i}}}=3.27$, $4.23$ and $7.51\,$meV at temperatures
of $T=1.6$ and $10\,$K. Data are also shown from a previously reported
experiment on polycrystalline LiFeAs, $T_{\mathrm{{c}}}=17\,$K, performed under similar
conditions on MERLIN~\cite{taylor_antiferromagnetic_2011}. In all
cases the scattering from a standard vanadium sample was used to normalize
the spectra and place them on an absolute intensity scale, with units
mb$\,$sr$^{-1}\,$meV$^{-1}\,$f.u.$^{-1}$, where $1\,$mb$=10^{-31}\,$m$^{2}$
and f.u. stands for formula unit of LaFePO, LaZnPO, Sr$_{2}$ScO$_{3}$FeP
or LiFeAs as appropriate.


\section{Results\label{sec:Results}}

\begin{figure}
\begin{centering}
\includegraphics[width=0.45\textwidth]{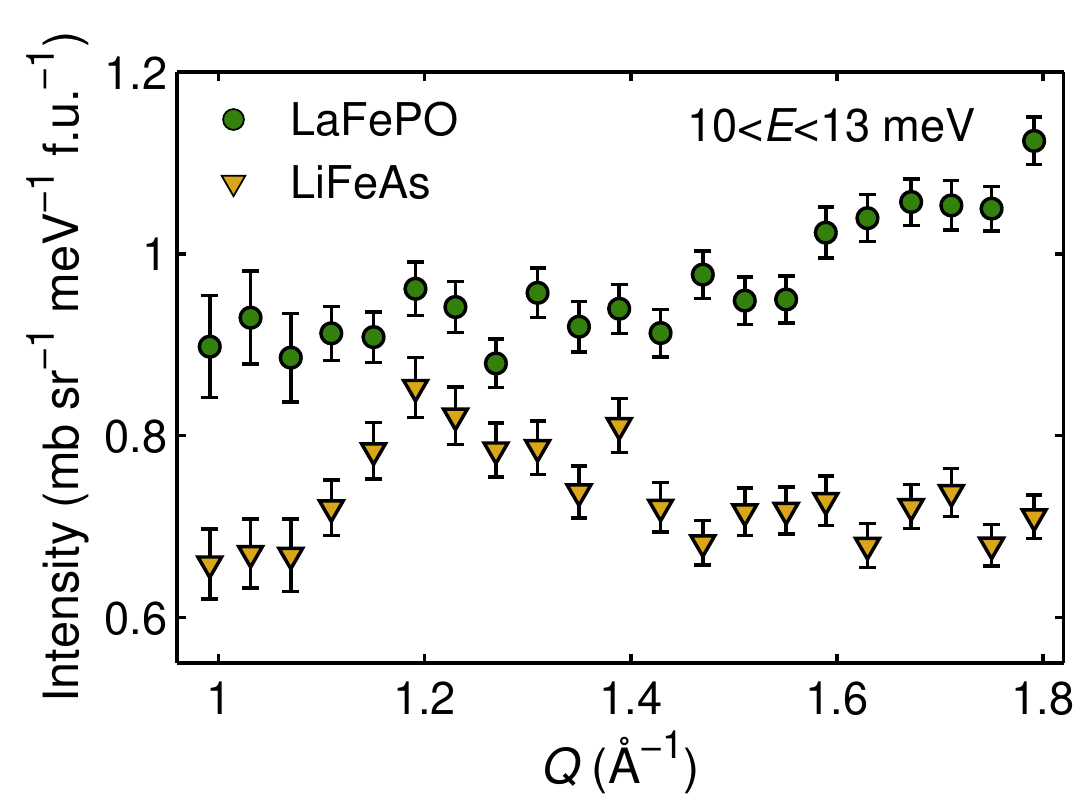}
\caption{\label{fig:LaFePO_vs_LiFeAs}  Constant-energy cuts showing the magnetic signal at $Q\approx1.2\,\mathrm{\AA}^{-1}$ in LiFeAs and
the same region in $(Q,E)$ space for LaFePO. Data were measured
on MERLIN at $T=20\,$K, with an incident energy $E_{\mathrm{{i}}}=25\,$meV,
and have been averaged over an energy range of 10--$13\,$meV.}
\end{centering}
\end{figure}

\begin{figure}
\begin{centering}
\includegraphics[width=0.45\textwidth]{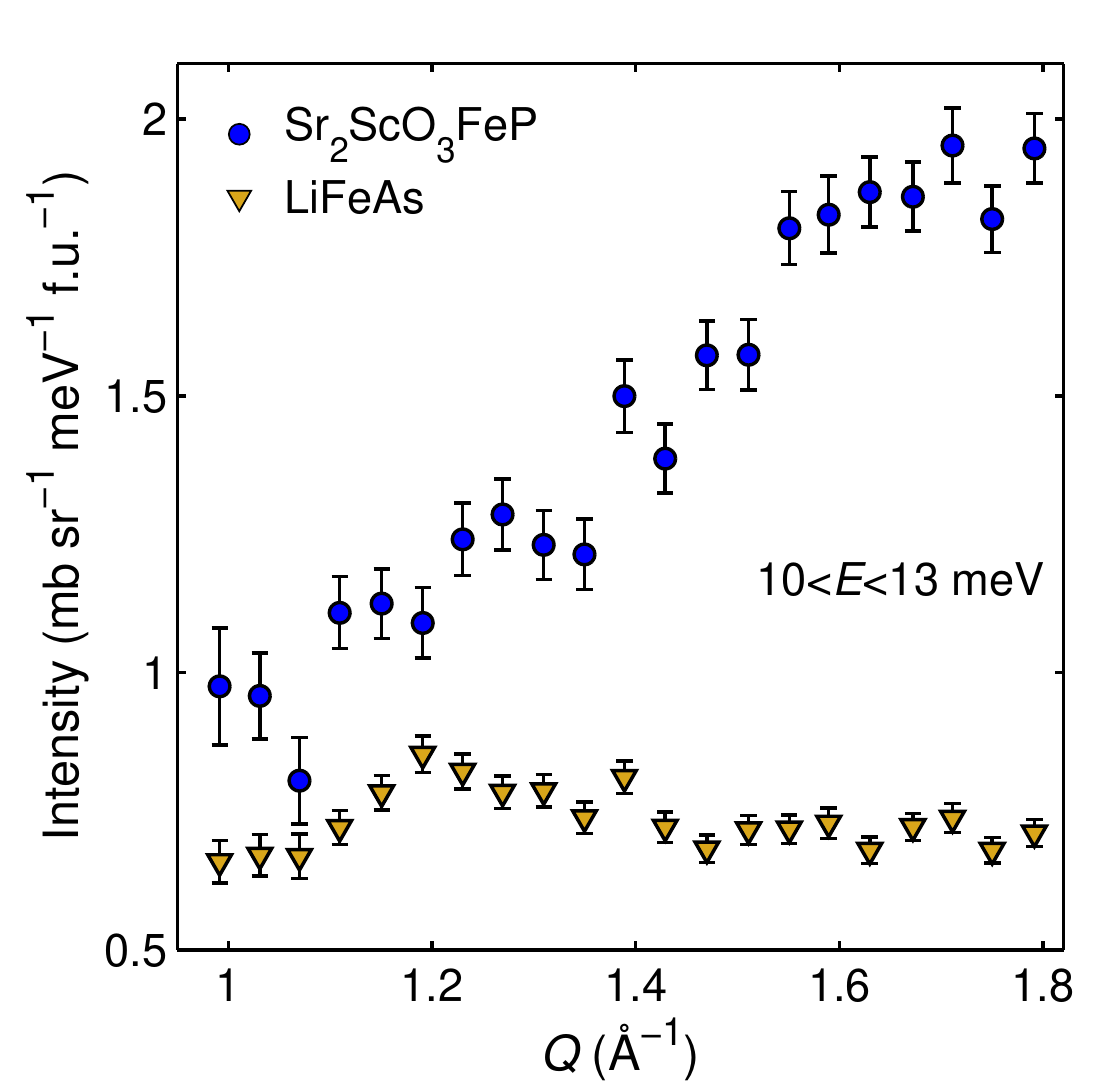}

\caption{\label{fig:Sr2ScO3FeP_vs_LiFeAs} Constant-energy cuts showing the
the magnetic signal at $Q\approx1.2\,\mathrm{\AA}^{-1}$ in LiFeAs
and the same region of $(Q,E)$ space for Sr$_{2}$ScO$_{3}$FeP.
Data were measured on MERLIN at $T=20\,$K, with an incident energy
$E_{\mathrm{{i}}}=25\,$meV, and have been averaged over an energy
range of 10--$13\,$meV. The Sr$_{2}$ScO$_{3}$FeP data have been
shifted down by 1 unit for clarity.}
\end{centering}
\end{figure}

Figures~\ref{fig:LaFePO_vs_LiFeAs} and~\ref{fig:Sr2ScO3FeP_vs_LiFeAs} compare the neutron inelastic scattering response of the phosphide materials with that of LiFeAs. All data in these figures are at temperatures greater than $T_{\mathrm{{c}}}$, to ensure that we are comparing samples in the normal state.
In figure~\ref{fig:LaFePO_vs_LiFeAs}, we show the comparison between constant-energy cuts from LaFePO ($T_{\mathrm{c}}\approx4.5\,$K)
and LiFeAs ($T_{\mathrm{c}}=17\,$K) data sets recorded at a temperature of 20$\,$K. The peak centered at $Q\approx1.2\,\mathrm{\AA}^{-1}$ in the LiFeAs data originates from quasi-2D spin fluctuations with characteristic in-plane wave vector close to ${\bf Q}_{\rm SDW}=\left(0.5,0\right)$~\cite{taylor_antiferromagnetic_2011, wang_antiferromagnetic_2011, qureshi_inelastic_2012}.  No such peak is present in the LaFePO data, whose intensity increases smoothly with $Q$ due to phonon scattering. By attempting to fit a Gaussian peak at $Q\approx1.2\,\mathrm{\AA}^{-1}$  to the LaFePO data we put an upper limit of about 15\% on the size of any such peak relative to the LiFeAs peak. To do this we fitted the LiFeAs data to a Gaussian peak on a linear background, we then constrained the peak width and centre and fitted the LaFePO data to the same function. We found a peak amplitude consistent with zero, with a fitting error 15\% of the size of the LiFeAs peak amplitude. As the spectra are normalized to one f.u., and one f.u. contains one Fe atom in both materials, a magnetic signal of given intensity per Fe would appear the same size in both spectra. Therefore, assuming the spin fluctuations have the same character in both materials, the spin fluctuations are at least 7 times weaker in LaFePO than in LiFeAs.


Figure~\ref{fig:Sr2ScO3FeP_vs_LiFeAs} shows the same LiFeAs data,
this time compared to a constant-energy cut from the Sr$_{2}$ScO$_{3}$FeP ($T_{\mathrm{c}}=15\,$K)
data measured at 20$\,$K. The Sr$_{2}$ScO$_{3}$FeP data have been
shifted down by one unit to aid comparison. The non-magnetic signal appears larger for Sr$_{2}$ScO$_{3}$FeP than LiFeAs because the spectra are normalized to one f.u., and the f.u. of Sr$_{2}$ScO$_{3}$FeP contains more atoms than that of LiFeAs. By performing a similar analysis as for LaFePO, we put an upper limit of 25\% on the size of any LiFeAs-type magnetic peak in the Sr$_{2}$ScO$_{3}$FeP data. 

\begin{figure}
\begin{centering}
\includegraphics[width=0.5\textwidth]{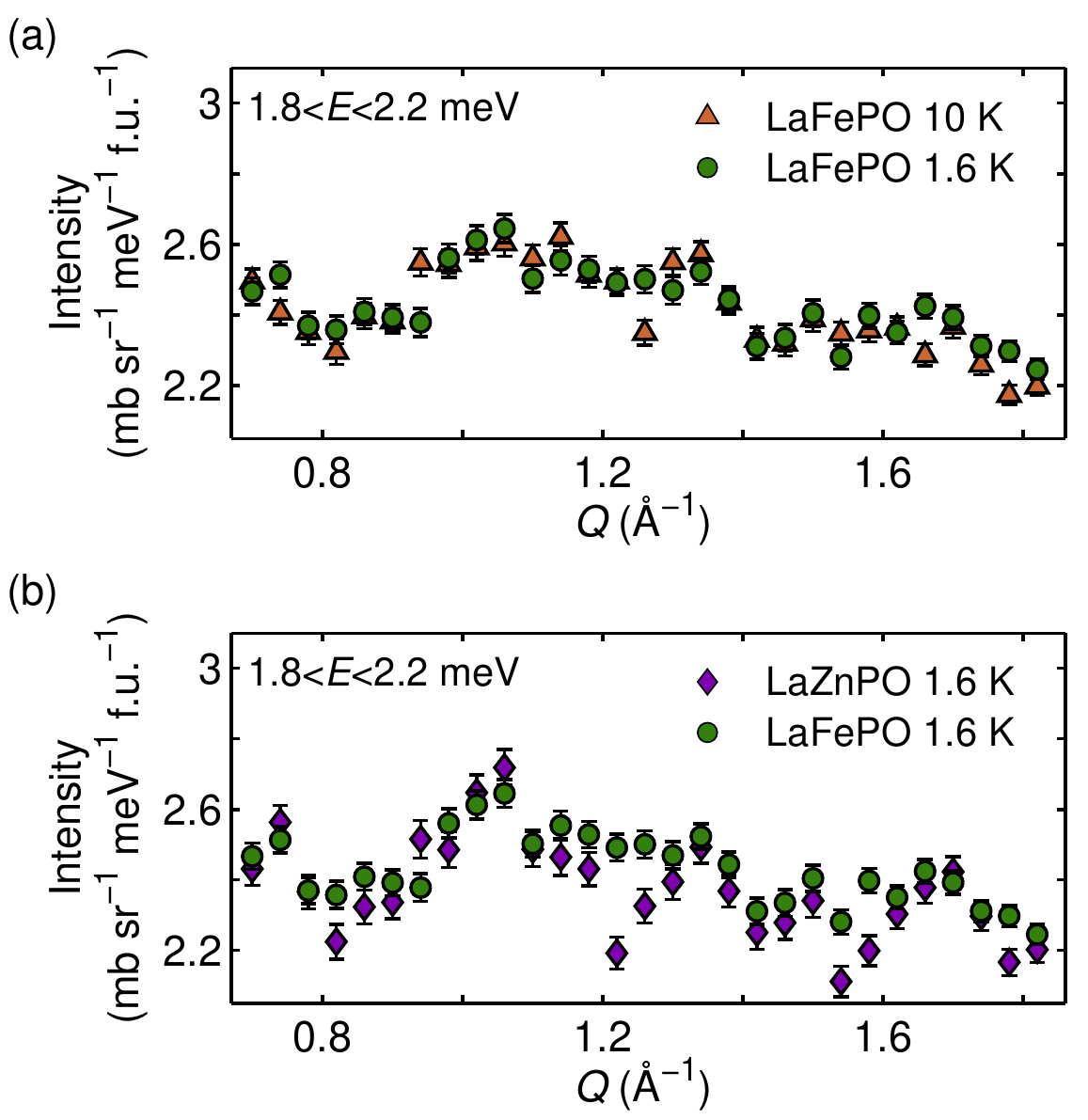}

\caption{\label{fig:LaFePO_Qcuts} Constant-energy cuts through data measured
on IN5 with an incident energy $E_{\mathrm{{i}}}=3.27\,$meV, averaged
over an energy range of 1.8--$2.2\,$meV. (a) Cuts through LaFePO data
taken at $10\,$K and $1.6\,$K, above and below $T_{\mathrm{c}}$ respectively.
(b) The same LaFePO data as in (a) measured at $1.6\,$K, with data
from LaZnPO also at $1.6\,$K.}
\end{centering}
\end{figure}

Figure~\ref{fig:LaFePO_Qcuts}(a) shows constant-energy cuts through data on LaFePO averaged over 1.8 to $2.2\,$meV, covering the energy where we would expect a magnetic resonance assuming the scaling relation $E_{\mathrm{res}}\approx5k_{\mathrm{B}}T_{\mathrm{c}}$. Data are shown from measurements at temperatures of 10$\,$K and 1.6$\,$K, i.e. above and below $T_{\mathrm{c}}$. The peak centered at $Q\approx1.05\,\mathrm{\AA}^{-1}$ is a feature of the non-magnetic background. This is shown in figure~\ref{fig:LaFePO_Qcuts}(b), which contains the same LaFePO 1.6$\,$K data together with an equivalent cut taken from the data on non-magnetic reference sample LaZnPO. All the features in the LaFePO data are reproduced in the LaZnPO data. If a spin resonance was present in figure~\ref{fig:LaFePO_Qcuts}(a), we would expect it to appear as an enhancement in intensity at $T<T_{\mathrm{c}}$ centred on the nesting wave vector of the Fermi surface of LaFePO~\cite{coldea_fermi_2008}, which corresponds to $Q\approx1.2\,\mathrm{\AA}^{-1}$. No such enhancement is found.

To be more quantitative, we made cuts like those in figure~\ref{fig:LaFePO_Qcuts}(a) through the data at
each of the neutron incident energies measured on IN5, and subtracted the 10\,K data from
the 1.6\,K data. We attempted to fit Gaussian peaks centred on $Q=1.2\,\mathrm{\AA}^{-1}$ to the subtracted data.  From the fit, we estimate the maximum area of any such peak to be 15\% of the peak at the resonance energy of LiFeAs.

\begin{figure}
\begin{centering}
\includegraphics[width=0.45\columnwidth]{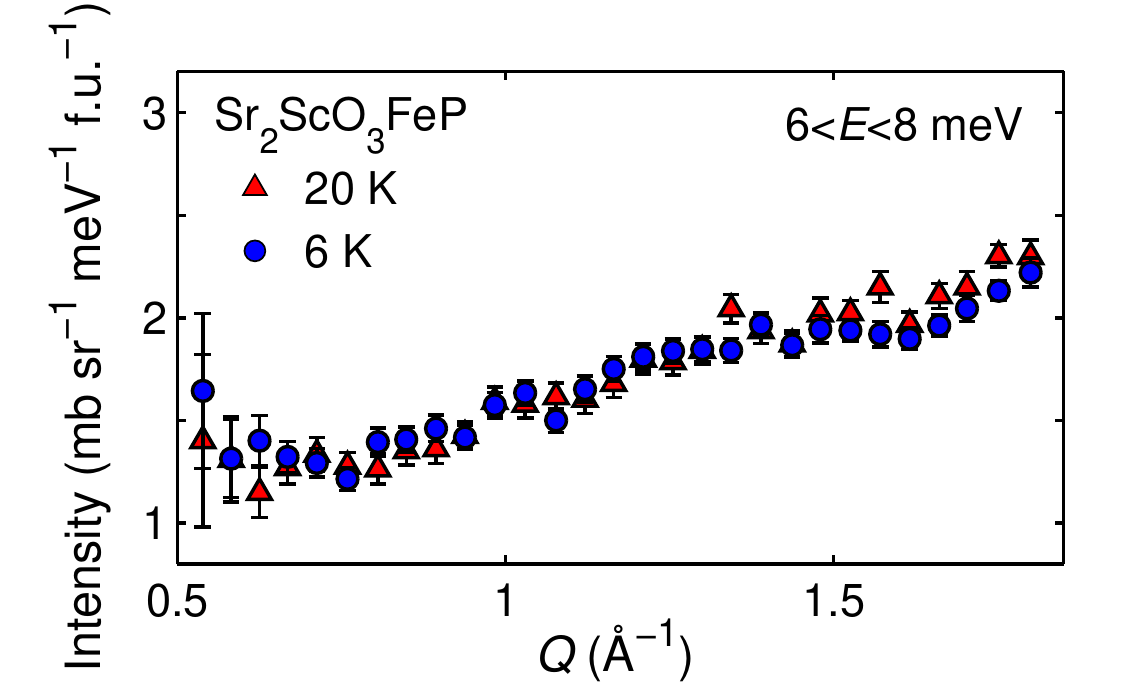}

\caption{\label{fig:Sr2ScO3FeP_Qcuts} Constant-energy cuts through Sr$_{2}$ScO$_{3}$FeP
data measured at 6\,K and 20\,K on MERLIN with an incident energy
$E_{\mathrm{{i}}}=25\,$meV, averaged over an energy range of 6--$8\,$meV.}
\end{centering}
\end{figure}

Figure~\ref{fig:Sr2ScO3FeP_Qcuts} shows wave vector cuts averaged over the energy range from 6 to 8$\,$meV, recorded at 20\,K and 6\,K. This is the approximate energy at which a spin resonance is expected in Fe-based superconductors with the same $T_{\mathrm{c}}$ as Sr$_{2}$ScO$_{3}$FeP. To within the statistical error, there is no difference between these data sets from above and below $T_{\mathrm{c}}$.

%

In addition to the selected cuts shown in figure~\ref{fig:LaFePO_Qcuts} and~\ref{fig:Sr2ScO3FeP_Qcuts},
the LaFePO and Sr$_2$ScO$_3$FeP spectra were examined at all  energies accessible in our measurements.  We compared runs recorded at different temperatures, as well as (in the case of LaFePO) the samples with and without Fe.  Energy
cuts at constant wave vector were also inspected in a similar way. No signal attributable
to magnetic fluctuations could be found for either material, neither in the superconducting nor in the normal state.


\section{Discussion}

The goal of this work was to determine whether prominent magnetic
fluctuations similar to those observed in other iron-based
superconductors are present in the iron phosphide systems LaFePO and
Sr$_{2}$ScO$_{3}$FeP. To within the limits of our experimental sensitivity,
we find no evidence for any magnetic fluctuations in these systems. Either the magnetic fluctuations are weaker than in other iron-based superconductors, or their characteristic length and time scales are too short to allow detection by inelastic neutron scattering.

This null result is interesting because there are reasonable expectations that magnetic fluctuations might exist in the iron phosphides. Firstly, like many of the iron arsenide and chalcogenide superconductors, the two iron phosphides studied here have Fermi surfaces with quasi-nested hole and electron pockets centred on $(0,0)$ and $(0.5,0)$, respectively~\cite{lu_electronic_2008,coldea_fermi_2008, shein_structural_2009, nakamura_highly_2010}. In the arsenides and chalcogenides, this quasi-nesting is widely thought to drive the SDW transition and explain the presence of strong spin fluctuations~\cite{mazin_unconventional_2008, johnston_puzzle_2010, stewart_superconductivity_2011, dai_magnetism_2012}.
Secondly, despite the evidence for nodal superconductivity in these phosphides~\cite{hicks_evidence_2009,fletcher_evidence_2009,sutherland_low-energy_2012,yates_evidence_2010}, magnetic fluctuations and a magnetic resonance are still expected to be of a typical strength for iron-based materials, as shown for example by the calculations of Maier {\it et al.}~\cite{maier_neutron_2009} for nodal and nodeless $s$-wave gaps. Moreover, magnetic fluctuations and a superconductivity-induced magnetic resonance were observed at $Q \approx 1.2\,$\AA$^{-1}$ by neutron scattering from the nodal superconductor BaFe$_2$(As$_{0.65}$P$_{0.35}$)$_2$~\cite{ishikado_s_like_2011}.
Thirdly, in the case of LaFePO, there is experimental evidence of magnetism. Anomalous static magnetic correlations were found in $\mu$SR measurements~\cite{carlo_static_2009}, and an NMR study on Ca-doped LaFePO (Ca doping increases the $T_{\mathrm{c}}$ of LaFePO to 7\,K) reported evidence for moderate ferromagnetic fluctuations~\cite{nakai_spin_2008, julien_enhanced_2008}. Ferromagnetic fluctuations would produce a peak in the neutron spectrum at $Q=0$. If the fluctuations were two-dimensional in character, this peak would extend to $Q>0$ with decreasing intensity; if the fluctuations were three-dimensional they would result in additional peaks at $Q_{(001)} =0.74\,\mathrm{\AA}^{-1}$,  $Q_{(002)} = 1.48\,\mathrm{\AA}^{-1}$, etc. No magnetic signal is observed in our data above the background at these or at any other wave vector probed, see figure~\ref{fig:LaFePO_Qcuts}.

 Because our data are normalized in absolute units we have been able to constrain the size of any magnetic signal in the phosphides. We have used LiFeAs as a reference. Despite poor Fermi surface nesting and no SDW order, the magnetic fluctuations observed at $Q\approx Q_{\rm SDW}$ in LiFeAs are of a typical strength for iron-arsenide-based superconductors~\cite{taylor_antiferromagnetic_2011,wang_antiferromagnetic_2011,ewings_high-energy_2008,christianson_unconventional_2008}. Our analysis has shown that a magnetic peak of the type found in the normal state of LiFeAs, if present, would have to be at least a factor of 4 (Sr$_2$ScO$_3$FeP) or 7 (LaFePO) smaller than in  other iron-based systems.

The absence of observable magnetic fluctuations in LaFePO is consistent with evidence that it is more metallic than iron arsenide superconductors~ \cite{kamihara_electromagnetic_2008}, and suggests that electronic correlations in LaFePO are weaker than in iron arsenides as reported in some previous studies~ \cite{si_strong_2008,qazilbash_electronic_2009}. Others, however, suggest that the correlations are of similar strength in the two families~\cite{coldea_fermi_2008, skornyakov_lda+dmft_2010}. In the iron arsenide systems a significant suppression of magnetic fluctuations has been observed in electron over-doped samples of  Ba(Fe$_{1-x}$Co$_{x}$)$_{2}$As$_{2}$~\cite{matan_doping_2010} and LaFeAsO$_{1-x}$F$_{x}$ ~\cite{wakimoto_degradation_2010}. However, $T_{\mathrm{c}}$ increases on electron doping in LaFePO$_{1-x}$F$_{x}$~\cite{kamihara_electromagnetic_2008}, which implies that LaFePO is not an intrinsically electron overdoped material.  It is possible, therefore, that the suppression of magnetic fluctuations is intrinsic to LaFePO and would be reproduced across its entire phase diagram, with no SDW phase proximate to superconductivity. In such a scenario, superconductivity in LaFePO could be controlled not by spin fluctuations but by a different pairing instability, as has been suggested by Thomale {\it et al.}~\cite{thomale_mechanism_2011}.

If spin fluctuations are involved in the pairing interaction then materials with weaker magnetic correlations would be expected to have lower $T_{\rm c}$ values. Our results for LaFePO are compatible with this expectation, but those for Sr$_2$ScO$_3$FeP are not.
The latter system has a relatively high maximum reported $T_{\rm c}$ of 17\,K, yet our results  lack the strong spin fluctuations characteristic of iron arsenide and chalcogenide superconductors with comparable $T_{\rm c}$ values. As discussed above, it appears that stoichiometric Sr$_2$ScO$_3$FeP is not a bulk superconductor, however our results along with the report by Ogino {\it et al.}~\cite{ogino_superconductivity_2009} suggest that Sr$_2$ScO$_3$FeP is close to the superconducting instability. We would therefore expect to see strong magnetic fluctuations as a precursor to the superconducting state. The absence of strong magnetic fluctuations in stoichiometric Sr$_2$ScO$_3$FeP could imply that superconductivity is not associated with a magnetic instability.

\section{Conclusions}

We have searched for, but did not find, a signal from magnetic fluctuations in the inelastic neutron scattering spectrum of two different iron phosphide materials. Magnetic fluctuations, if present, are significantly weaker than in iron arsenide and chalcogenide systems. This suggests that magnetic fluctuations might not play as significant a role in iron phosphide superconductors as they do in other iron-based superconductors. Because of its relatively high superconducting transition temperature ($T_{\mathrm{c}} = 17$\,K), identification of the precise composition of the superconducting phase in Sr$_2$ScO$_3$FeP would be a significant step towards an understanding of the mechanism of superconductivity in the iron phosphide superconductors.

\ack
This work was supported by the UK Engineering \&
Physical Sciences Research Council and the Science \& Technology Facilities Council. We thank A. J. Corkett and M. J. Pitcher for help with synthesis, J. R. Stewart for help with the neutron scattering experiments, and R. Thomale for comments on the manuscript.

\section*{References}
\bibliographystyle{AET_bst}
\bibliography{Phosphides}

\end{document}